\documentstyle[prl,aps]{revtex}
\input epsf.tex
\twocolumn

\def\eqn{\begin{equation}}
\def\endeqn{\end{equation}}
\def\eqna{\begin{eqnarray}}
\def\endeqna{\end{eqnarray}}
\begin{document}
\draft
\title
{
\rightline{\bf CLNS98/1598}
Sixth-Order Vacuum-Polarization Contribution\\ to the Lamb Shift  
of the Muonic Hydrogen}

\author{
T. Kinoshita\thanks{e-mail: tk@hepth.cornell.edu}
 }

\address{ Newman Laboratory of Nuclear Studies,
Cornell University, Ithaca, NY 14853 }

\author{
M. Nio\thanks{e-mail: makiko@phys.nara-wu.ac.jp}
 }

\address{ Graduate School of Human Culture,
Nara Women's University, Nara, Japan 630 }

\date{\today}
\maketitle

\begin{abstract}
The sixth-order
electron-loop vacuum-polarization contribution to
the $2P_{1/2} - 2S_{1/2}$ Lamb shift of 
the muonic hydrogen ($\mu^{-} p^+$ bound state)
is evaluated numerically.
Our result is
0.007608~(1) meV. 
This eliminates the largest theoretical uncertainty.
Combined with
the proposed precision measurement of the Lamb shift it will lead to
a precise determination of the
proton charge radius.
\end{abstract}


The muonic hydrogen, the $\mu^- p^+$ bound state,
differs from the ordinary hydrogen atom in two important respects.
One is that the vacuum-polarization effect is much more
important than other radiative corrections.
The other is that it is more sensitive to the hadronic
structure of the proton.
Thus it provides a means of testing aspects of QED
significantly different from those of the hydrogen atom.

The muonic hydrogen 
has a long-lived $2S$ meta-stable state.
This makes it possible to measure the $2P_{1/2}-2S_{1/2}$ 
Lamb shift to about 10 ppm 
level using the phase-space compressed muon beam technique \cite{taqqu}. 
At present, however,
theoretical precision is limited
to about 50 ppm. This uncertainty comes mainly from  
the unknown contribution $\Delta E^{(6)}$
of the sixth-order electron vacuum-polarization effect
\cite{pachucki}.

In this  paper we report the result of our 
evaluation of $\Delta E^{(6)}$.
Our result is
\eqna
\Delta E^{(6)}&=& 
                 0.120~045~(12)                        
~m_r (Z\alpha)^2 \left ( {\alpha \over \pi} \right )^3
\nonumber \\
               &=& 0.007~608~(1) ~~\mbox{meV},
\label{ourresult}
\endeqna
where $Z= 1$ for the proton and 
$m_r$ is the reduced mass of the $\mu^- p^+$ system:\cite{const}
\eqna
     m_r & = &{ m_{\mu} m_p \over m_{\mu} + m_p } = 94.964~485~(28)~\mbox{MeV}, 
\nonumber \\
     m_{\mu}&=& 105.658~389(34)~~\mbox{MeV},  
\nonumber \\
     m_p&=&938.272~31(28)~~\mbox{MeV}~~. 
\endeqna     

We have also evaluated the main part of $\Delta E^{(6)}$ using the
Pad\'{e} approximation of vacuum-polarization function 
\cite{BB}.
The result (\ref{Ep6Pade}) 
is in good agreement with 
the direct calculation (\ref{Ep6}).

The contribution to
the $2P_{1/2}-2S_{1/2}$ Lamb Shift of the muonic hydrogen  
due to the effect of the
electron-loop vacuum-polarization on a single Coulomb photon
can be expressed as an integral over the vacuum-polarization
function $\Pi (q^2 )$.
Here $q$ may be either space-like or time-like.
The first choice ($q^2 < 0$) leads to the integral
\eqn
\Delta^{(I)} E= \int{d^3q \over (2\pi)^3} \tilde{\rho}(a^2) 
               { -4\pi Z\alpha \over \vec{q}^2}
             \left [    -\Pi(-\vec{q}^2) \right ] ~. 
\label{lsRe}
\endeqn
%
%
Here $\tilde{\rho}$ is equal to
$\tilde{\rho}_{2P}-\tilde{\rho}_{2S}$, 
$\tilde{\rho}_{2P}$ 
and $\tilde{\rho}_{2S}$ 
being Fourier 
transforms of squares of
non-relativistic Coulomb wave functions 
for the $2P$ and $2S$ states: 
\eqn
\tilde{\rho}_{2P(2S)}= \int d^3r  |\phi_{2P(2S)}(\vec{r})|^2 
e^{-i \vec{q}\cdot\vec{r}} .
\label{rho2ps}
\endeqn
Carrying out the integration we obtain
\eqn 
\tilde{\rho}_{2P}= { 1-a^2 \over (1 + a^2)^4 }~,
~~~~\tilde{\rho}_{2S}= {1 - 3a^2 + 2a^4  \over (1 + a^2)^4 }~,
\endeqn
where $ a = |\vec{q} |/(Z\alpha m_r)$ and $\tilde{\rho}_{2P}$ is 
averaged over three degenerate states.

The second choice ($q^2 > 0$) gives rise to the integral \cite{pachucki} 
\eqn
\Delta^{(II)} E = m_r (Z\alpha)^2 
\int_4^{\infty}  dt 
u(t) { \beta^2  \over 2 (1 + \beta \sqrt{t} )^4} ~,
\label{lsIm}
\endeqn      
where
\eqn
\beta = { m_e \over m_r \alpha} = 0.737~383~76~(30)
\endeqn
and 
\eqn
u(t)= {1 \over \pi} Im \Pi( q^2 = t m_e^2) ~~.
\endeqn 

Although Eqs. (\ref{lsRe}) and (\ref{lsIm})
are analytically equivalent,
they are totally different
from the viewpoint of numerical integration.
Thus they provide a useful check whenever both real and imaginary 
parts of $\Pi$ are available.
For diagrams containing several vacuum-polarization loops
in one Coulomb photon line,
Eqs. (\ref{lsRe}) and (\ref{lsIm}) must be modified accordingly.
Insertion of vacuum polarization loops in several
Coulomb photon lines can be handled by the non-relativistic bound-state
perturbation theory.

\begin{figure}
\centerline{\epsfbox{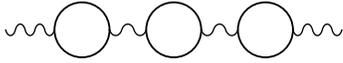}}
\vspace{0.2cm}
\caption{
Three second-order vacuum-polarization diagrams
inserted in the Coulomb photon line
exchanged by the muon and the proton.
}
\label{figlsp2:3}
\end{figure}

Let us first consider insertion of three second-order
vacuum-polarizations 
in a Coulomb photon (see Fig.\ref{figlsp2:3}). 
The contribution $\Pi^{(p2:3)}(q^2)$ 
of this improper diagram
can be expressed in terms of the second-order vacuum-polarization
function $\Pi^{(2)}(q^2)$ as
\eqn
\Pi^{(p2:3)}(q^2)=  (\Pi^{(2)}(q^2))^3~,
\endeqn
where $\Pi^{(2)}$ is known analytically and 
has the spectral function 
\eqn 
u^{(2)}(t)={1 \over 3} {\alpha \over \pi} \sqrt{1 - {4 m_e^2 \over q^2}}
\left ( 1 + {2m_e^2 \over q^2} \right ) ,
~~~q^2 \geq 4 m_e^2 .
\endeqn
Substituting $\Pi^{(p2:3)}$ in Eq. (\ref{lsRe}) and 
evaluating the integral numerically,\footnote{This and subsequent 
integrals are evaluated numerically 
either on DEC$\alpha$ or 
on Fujitsu-VX of NWU, or on both,
by the adaptive-iterative Monte-Carlo 
subroutine VEGAS \cite{lepage}.}
we find
\eqn
\Delta^{(I)} E^{(p2:3)}=
0.006~253~4~(6)
~m_r (Z \alpha)^2 \left (
{\alpha \over \pi }\right )^3  ~.
\label{Ep2:3}
\endeqn
%
%
%
The result of the second method 
(\ref{lsIm}) agrees with (\ref{Ep2:3}):
\eqn
\Delta^{(II)} E^{(p2:3)}=
0.006~253~9~(10)
~m_r (Z \alpha)^2 \left (
{\alpha \over \pi }\right )^3 ~.
\label{ImEp2:3}
\endeqn
%
%
Another evaluation of $\Delta^{(I)} E^{(p2:3)}$ 
using the parametric-integral form of $\Pi^{(2)}$ 
given in Ref. \cite{KL2} leads to
\eqn
\Delta^{(I)} E^{(p2:3)}=
0.006~253~8~(8)
~m_r (Z \alpha)^2 \left (
{\alpha \over \pi }\right )^3 ~.
\label{Ep2:3param}
\endeqn


\begin{figure}
\centerline{\epsfbox{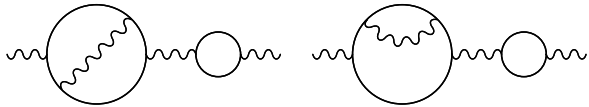}}
\vspace{0.2cm}
\caption{Insertion of one
second- and one fourth-order vacuum-polarization diagrams
in the Coulomb photon line exchanged by the muon and the proton.}
\label{figlsp4p2}
\end{figure}

The next contribution  comes from 
diagrams involving one second-order
and one fourth-order vacuum-polarization insertions
(see Fig. \ref{figlsp4p2}). 
This contribution is given in terms of
\eqn
\Pi^{(p4p2)}(q^2)= -2\Pi^{(2)}(q^2) \Pi^{(4)}(q^2)~,
\endeqn
where $\Pi^{(4)}$ is 
the fourth-order vacuum-polarization function \cite{kallen}.
Substituting $\Pi^{(4)}$ 
into Eqs. (\ref{lsRe}) and (\ref{lsIm}) we obtain
\eqn
\Delta^{(I)} E^{(p4p2)}=
0.046~248~(5)
~m_r (Z\alpha)^2 \left ( {\alpha \over \pi }\right )^3 ~,
\label{Ep4p2}
\endeqn
and
\eqn
\Delta^{(II)} E^{(p4p2)}=
0.046~243~(16)
~m_r (Z\alpha)^2 \left ( {\alpha \over \pi }\right )^3 ~.
\label{ImEp4p2}
\endeqn
%
We also evaluated $\Delta^{(I)} E^{(p4p2)}$ 
using the parametric-integral form of $\Pi^{(4)}$ \cite{KL2}:
\eqn
\Delta^{(I)} E^{(p4p2)}=
0.046~250~(2)
~m_r (Z \alpha)^2 \left (
{\alpha \over \pi }\right )^3 ~.
\label{Ep4p2param}
\endeqn
%
%

\begin{figure}
\centerline{\epsfbox{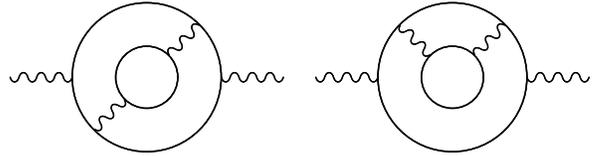}}
\vspace{0.2cm}
\caption{Sixth-order vacuum-polarization diagrams
with a second-order vacuum-polarization inserted
in the fourth-order vacuum-polarization diagrams. }
\label{figlsp4(p2)}
\end{figure}

The third contribution comes from the sixth-order vacuum-polarization
term $\Pi^{(p4(p2))}$ 
obtained by inserting a second-order vacuum-polarization loop in the
fourth-order vacuum-polarization diagram (see Fig.\ref{figlsp4(p2)}).
The form of  $\Pi^{(p4(p2))}$ 
convenient for numerical integration
is as integral over Feynman parameters \cite{KL2}.
This can be done  easily by adapting to the Lamb shift
the program written previously for
the electron $g-2$
\cite{KL1}. 
This leads to
\eqn
\Delta^{(I)} E^{(p4(p2))}=
0.013~628~(6)
~m_r (Z\alpha)^2 \left ({\alpha \over \pi }\right)^3 ~.
\label{Ep4(p2)}
\endeqn
The $\overline{\rm MS}$ renormalized imaginary part of $\Pi^{(p4(p2))}$ 
is known in a two dimensional integral form \cite{chkst}. 
Converting it to the on-shell renormalized one 
and using Eq. (\ref{lsIm}), we obtained
\eqn 
\Delta^{(II)} E^{(p4(p2))}=
0.013~626~(1)
~m_r (Z\alpha)^2 \left ({\alpha \over \pi }\right)^3 ~.
\label{ImEp4(p2)}
\endeqn


\begin{figure}
\centerline{\epsfbox{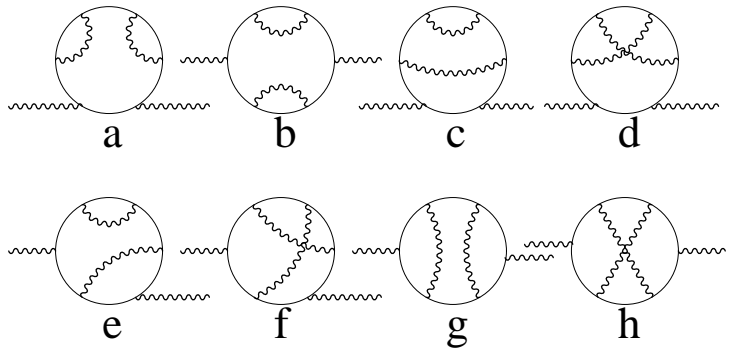}}
\vspace{0.2cm}
\caption{Sixth-order vacuum-polarization diagrams
with a single electron loop.}
\label{figlsp6}
\end{figure}

%
The fourth contribution 
comes from the sixth-order vacuum-polarization
diagrams with a single electron loop. 
The exact form of this contribution
is known only in a parametric-integral form \cite{KL2}.
Its imaginary part is not available in
a form convenient for numerical work.
We have therefore evaluated it using Eq. (\ref{lsRe}) only.
There are eight topologically distinct diagrams 
(see Fig.\ref{figlsp6}).
Each diagram can be written as a sum of various
divergent terms and a finite part $\Delta \Pi^{(6i)}$,
where $i = a, b, ..., h$.
After renormalization the sum of these diagrams is
free from any divergence and
can be written as \cite{KL2}
\eqna 
\Pi^{(p6)}&=&2(\Delta\Pi^{(6a)}+  \Delta\Pi^{(6c)}
            +\Delta\Pi^{(6d)}+  \Delta\Pi^{(6f)} )
\nonumber \\
          &+&\Delta\Pi^{(6b)}+ 4\Delta\Pi^{(6e)}
            +\Delta\Pi^{(6g)}+  \Delta\Pi^{(6h)}
\nonumber \\
            &-&4\Delta B_2 \Pi^{(4)} 
-2[\Delta B_{4a} + \Delta L_{4x}+2\Delta L_{4c}
\nonumber \\
&+&\Delta B_{4b}
+\Delta L_{4l}+2\Delta L_{4s} + { 3 \over 2}(\Delta B_2)^2 ] \Pi^{(2)}
\nonumber \\
    &-&2(\Delta\delta m_{4a} + \Delta \delta m_{4b} ) \Pi^{(2\ast)}~,
\endeqna
where $\Delta B_2,~ \cdots$, are finite parts of
renormalization constants
and $\Pi^{(2)}$ and $\Pi^{(4)}$
are renormalized vacuum-polarization functions of
second- and fourth-order, respectively. 
$\Pi^{(2\ast)}$ is the second-order vacuum-polarization
function with a mass insertion vertex.
Precise definitions of these functions
are given in Ref. \cite{CK}.
The numerical values of the coefficients of $\Pi^{(4)}$, $\Pi^{(2)}$
and $\Pi^{(2\ast)}$ are 
\eqna
\Delta B_2 &=& {3 \over 4}~{\alpha \over \pi}~, 
\nonumber \\
\Delta B_{4a}+ ~\cdots~ +{3\over 2}(\Delta B_2)^2 &=&  
  0.871~680~(27)
  \biggl({\alpha \over \pi}\biggr )^2 ,
\nonumber \\
\Delta \delta m_{4a} +\Delta \delta m_{4b} &=&  
  1.906~340~(21)  
  \biggl({\alpha \over \pi}\biggr )^2 ~ ,
\endeqna
where the last two are new evaluations.
The Lamb Shift contributions from $\Pi^{(4)}$, $\Pi^{(2)}$, and $\Pi^{(2\ast)}$
can be easily obtained by numerical integration:
\eqna
\Delta E^{(p4)}&=& 
0.045~922~7~(4)
~m_r (Z\alpha)^2 \biggl ({\alpha \over \pi}\biggr )^2~ ,
\nonumber \\
\Delta E^{(p2)}&=& 
0.017~452~8~(3)
~m_r (Z\alpha)^2 { \alpha \over \pi }~,
\nonumber \\
\Delta E^{(p2\ast)}&=& 
-0.009~001~8~(2) 
~m_r (Z\alpha)^2 { \alpha \over \pi }.
\endeqna

The Lamb Shift contributions $\Delta E^{(p6a)}, \cdots$, coming from 
the ultraviolet- and infrared-finite parts of diagrams
$\Delta \Pi^{(6a)}, \cdots$, are numerically evaluated.
The results are summarized in Table I. 
The second  and third columns list the results of integration carried out
in double precision
and quadruple precision, respectively.
The purpose of the latter calculation is to see whether 
the former indicates sign of losing
significant digits due to
rounding-off, which we call $digit$-$deficiency$ 
problem and is the major source of
uncertainty of {\it on-the-computer} renormalization \cite{ourBB}.
The excellent agreement between two calculations shows that
the estimated error of the former 
is not significantly affected by the $digit$-$deficiency$ problem and
can be safely assumed to be mostly statistical.
We therefore choose the double precision value,
which has higher statistics, as our best estimate:
\eqn
\Delta^{(I)} E^{(p6)}
=0.017~410~(9)
~m_r (Z\alpha)^2 \left ( {\alpha \over \pi }\right )^3 ~.
\label{Ep6}
\endeqn

%

%
As a cross-check, we also evaluated $\Delta E^{(p6)}$
using the Pad\'{e}-approximation of the vacuum-polarization function
from Ref. \cite{BB}.
We did this using both methods I and II.
The [2/3] and [3/2] Pad\'{e} approximations
give nearly identical results.
Taking their average  we obtain
\eqna
\Delta^{(I)} E_{Pad\acute{e}}^{(p6)}
&=&0.017~414~9~(25)
~m_r (Z\alpha)^2 \left ( {\alpha \over \pi }\right )^3 ,
\nonumber   \\
\Delta^{(II)} E_{Pad\acute{e}}^{(p6)}
&=&0.017~414~9~(26)
~m_r (Z\alpha)^2 \left ( {\alpha \over \pi }\right )^3 .
\label{Ep6Pade}
\endeqna
These results are consistent with each other
and agree with (\ref{Ep6}) to three significant digits,
or within one standard deviation of (\ref{Ep6}).
Obviously either (\ref{Ep6}) or (\ref{Ep6Pade})
has sufficient precision
as far as comparison with experiment is concerned.
Note, however, that the uncertainties given in (\ref{Ep6Pade})
are those resulting from numerical treatment of the Pad\'{e}
approximation and do not include those caused by the Pad\'{e} method
itself. 
It is argued in a separate paper \cite{ourBB}
that the uncertainty
of the Pad\'{e} model itself is about 0.001 percent
and hence the true value will be found well within the
uncertainties given in
(\ref{Ep6Pade}).

%
\begin{figure}
\centerline{\epsfbox{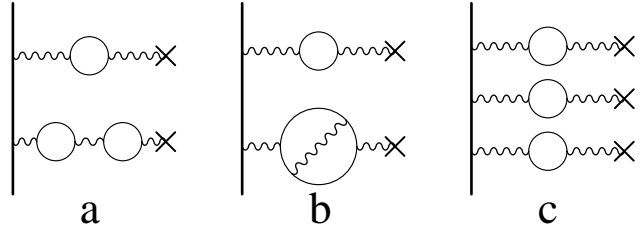}}
\vspace{0.2cm}
\caption{Representative sixth-order diagrams in which
vacuum-polarization insertion occurs
in two and three Coulomb photon lines.
Vertical lines represent the muon moving in the
Coulomb potential generated by the proton,
which is indicated by a $`` \times$".}
\label{lsp6fig5}
\end{figure}

Thus far we considered only diagrams in which one
Coulomb photon line is modified by the electron-loop
vacuum polarization.
Additional contributions of order $\alpha^3$ arise from
the diagrams of Fig. \ref{lsp6fig5} in which
two and three Coulomb photons are modified by 
vacuum polarization.
Their contributions to the Lamb shift
can be found by the bound-state perturbation theory:
\eqna
\Delta E ({\rm Fig.5a})
&=&0.009~166~(2)
~m_r (Z\alpha)^2 \left ( {\alpha \over \pi }\right )^3 ,
\nonumber   \\
\Delta E ({\rm Fig.5b})
&=&0.024~805~(3)
~m_r (Z\alpha)^2 \left ( {\alpha \over \pi }\right )^3 ,
\nonumber   \\
\Delta E ({\rm Fig.5c})
&=&0.002~535~(1)
~m_r (Z\alpha)^2 \left ( {\alpha \over \pi }\right )^3 .
\label{Ep6fig5}
\endeqna
In this calculation we used the reduced non-relativistic Coulomb
Green function for $2S$ and $2P$ states given by Eqs. (23) and (24)
of Ref. \cite{pachucki}.

Collecting  (\ref{Ep2:3}), (\ref{Ep4p2}), (\ref{Ep4(p2)}), 
(\ref{Ep6}), and (\ref{Ep6fig5}), 
we obtain the total contribution
to the Lamb shift (\ref{ourresult}) due to 
the sixth-order vacuum-polarization effect.

Evaluation of various lower-order contributions
to the $2P_{1/2} - 2S_{1/2}$
Lamb shift ${\cal L}$ of the muonic hydrogen 
are summarized in Ref. \cite{pachucki}\footnote{
K. Pachucki informed us that F. Kottman pointed out  
that the sum of all contributions listed in  Ref. \cite{pachucki}  
was 206.049 meV, not 205.932 meV.}. 
In addition we have obtained the hadronic vacuum-polarization
correction of 0.0113(3) meV
following Ref. \cite{friar}.
%
These results and our result (\ref{ourresult})
lead to the most precise theoretical prediction
\eqn
{\cal L} = (206.068~(2) - 5.197~5~r_p^2  )~ {\rm meV},
\label{lamb}
\endeqn
where $r_p$ is the proton charge radius in units of fm.
The uncertainty in the first term of (\ref{lamb})
is our estimate of theoretical error.

To improve the theoretical prediction further,
it is necessary to have better estimate of the effect
to the Lamb shift and hyperfine structure of the
muonic hydrogen due to the proton's internal structure
beyond elastic form factors.
Recently the proton polarizability correction to the
hyperfine structure of the hydrogen and muonic hydrogen
was obtained \cite{russian}.
There are also references 
for ordinary hydrogen and deuterium \cite{pachucki2,khrip}.
Unfortunately they are not directly applicable to the muonic
hydrogen because of very different energy scale.

Measurement of ${\cal L}$ to 10 ppm, or 0.002 meV,
will lead to improvement in the value of $r_p^2$
by an order of magnitude over those determined from the elastic
scattering form factor measurements,
making it possible to resolve 
the long-standing discrepancy between \cite{simon} and \cite{hand}.
The new value of $r_p^2$ will also play an important role in
testing the validity of QED in terms of high precision measurements
of the hydrogen atom \cite{udem}.
Another impact of
accurate determination of $r_p^2$ will be 
to stimulate  evaluation of $r_p^2$
from the lattice QCD more precise and reliable 
than those available at present \cite{leinweber}.


We thank D. Taqqu for communicating about
the proposed measurement of the Lamb shift of the muonic hydrogen.
We thank K. Pachucki for pointing out the need to include
the contribution of Fig. \ref{lsp6fig5}, for bringing our
attention to Ref. \cite{friar}.
We also thank K. Chetrykin, J. H. K\"{u}hn, R. Harlander, 
and M. Steinhauser for informing us of Ref. \cite{chkst}. 
The work of
T. K. is supported in part by the U. S. National Science Foundation.
The work of
M. N. is supported in part by the Grant-in-Aid (No. 10740123)
of the Ministry of Education, Science, and Culture, Japan.  

\begin{table}
\caption{Contributions 
to the $2P_{1/2}-2S_{1/2}$ muonic hydrogen Lamb shift from
the sixth-order vacuum  polarization diagrams
with a single electron loop.
The overall factor $m_r (Z\alpha)^2(\alpha/\pi)^3 $ is omitted.
The second and third columns give results of integration
in double precision and quadruple precision, respectively.
Their difference is listed in column 4.
\label{table2}
}
\begin{tabular}{crrr}
\hspace{0mm} {\rm Term } \hspace{2mm} 
        &\hspace{2mm} {\rm Doub.~precis. }         
        &\hspace{2mm} {\rm Quad.~precis. }       
        &\hspace{2mm} {\rm Difference }\hspace{0mm}         
\\ \hline
$\Delta E^{(6a)} $&  0.044~769~(4)  &  0.044~739~(51)   
&  0.000~030~(52)
\\ \hline
$\Delta E^{(6b)} $&  0.028~654~(4)  &  0.028~640~(35)   
&  0.000~014~(36)
\\ \hline
$\Delta E^{(6c)} $& -0.025~393~(3)  & -0.025~368~(23)   
& - 0.000~025~(24)
\\ \hline
$\Delta E^{(6d)} $& -0.026~376~(2)  & -0.026~371~(21)   
& - 0.000~005~(22)
\\ \hline
$\Delta E^{(6e)} $&  0.151~356~(4) &  0.151~334~(46)   
&  0.000~022~(47)
\\ \hline
$\Delta E^{(6f)} $& -0.067~139~(3)  & -0.067~144~(30)   
&  0.000~005~(31)
\\ \hline
$\Delta E^{(6g)} $&  0.019~536~(3) &  0.019~540~(23)   
& - 0.000~004~(24)
\\ \hline
$\Delta E^{(6h)} $&  0.025~877~(2)  &  0.025~858~(22)   
&  0.000~019~(23)
\\
\end{tabular}
\end{table}

  
%
%
%
%
%
%
\end{document}